| |
|---|
| ESO/Cou-1928<br>Date: 16.11.2020 |

EUROPEAN ORGANISATION FOR ASTRONOMICAL

RESEARCH IN THE SOUTHERN HEMISPHERE

---

| | |
|---|---|
| **Council**<br>155th Meeting, Videoconferencing<br>1-2 Dec 2020 | **For information** |

# A Report on the Impact of Satellite Constellations

This document is **ESO INTERNAL** until its review, afterwards it is for **PUBLIC DISTRIBUTION**

Distribution to Council members, their colleagues with a need-to-know, and their supervisors is authorised. This distribution also applies to AU Observers.

Council is invited to note this document.

## EXECUTIVE SUMMARY

Up to 100,000 satellites could be launched into Low Earth Orbit (LEO) in the coming decade. Assuming the two most advanced companies' plans are realised, close to 80,000 satellites will be present at a variety of altitudes between 328 – 1,325 km. At Paranal, more than 5,000 satellites will be over the horizon at any given time. Of these, depending on the hour of night and season, a few hundred to several thousand will be illuminated by the sun and potentially detectable. Satellites show a very strong concentration towards the local horizon, with over 50% of the satellites below 20 degrees elevation. This report informs Council of the impacts on ESO facilities, mitigation measures that ESO could adopt in the future, and the various community efforts in which ESO is involved.

ESO's current and future optical/infrared telescopes are affected for observations at twilight, particularly with low elevation and long exposure times. For the rest of the night, observations are affected only to a small degree. Even at twilight, the impact remains lower or at worst comparable to the % observing time loss typically expected for technical downtime, and significantly lower than weather losses, and is therefore considered manageable. Incorporating satellite locations into scheduling tools could reduce the number of affected frames, but would introduce additional constraints making the optimal scheduling of observations less efficient. While in a quantitative sense, the impacts are minimal, this analysis cannot account for the loss of unknown or serendipitous science as a result of affected frames or the possibility that observations that are time critical or that cannot be rescheduled could be required in an area of the sky heavily populated with satellites (e.g. optical counterparts of gravitational wave source or a gamma ray burst).

Future initiatives aiming to construct large, wide-field, optical telescopes will likely face substantial design constraints to deal with the impact of satellite constellations.

Only one of the leading satellite constellation operators is using ALMA frequencies (Band 1). Some observations in the affected portions of this band could require longer observing times to overcome the increased noise. While the other ALMA bands are not currently affected, constant vigilance is required in terms of monitoring radio frequency spectrum use requests from industry and government projects. ESO will work with other ALMA partners and representatives from the radio astronomy community to seek agreement with satellite operators and appropriate government regulators to ensure that radio quiet zones are maintained, and space-born radio interference is also considered.

While ESO facilities are not currently substantially affected, ESO will continue to deploy its expertise and resources to support the community in understanding the broader impacts, and in raising awareness amongst space industry and policymaker stakeholders. ESO will also maintain awareness of the possible effects as a result of strongly impacted facilities (e.g. the Vera C. Rubin Observatory), which could provide observation targets for ESO telescopes, or engage in joint campaigns in the future.

**Table of Contents**

# 1. Introduction

## 1.1 Purpose of report

Following the launch on 23 May 2019 of the first batch of SpaceX's *Starlink* satellite constellation, the astronomy community has become increasingly aware of the impacts that constellations will have on ground-based astronomy across all wavelengths. The high numbers of satellite units in planned constellations and their relatively low altitude means that a substantially higher number of bright and moving objects will be present in the night sky at any given location and particularly in and near twilight hours. Moreover, all of these satellite units will be transmitting in the radio frequency spectrum and observatories can no longer rely on the protections offered by localized radio quiet zones.

This report informs Council of recent developments and ESO actions, summarises the impacts on current and future ESO facilities, describes the mitigation measures that ESO could adopt in the future, and describes the various community efforts in which ESO is involved.

## 1.2 Background

The optical astronomy community was taken largely by surprise upon the launch of the first batch SpaceX's *Starlink* satellite constellation on 23 May 2019. The immediate post-launch configuration of the 60 satellite units were visible as a very bright 'string of pearls' travelling at high velocity across the night sky, and generated substantial public, media and astronomy community interest. ESO, along with many other observatories, national agencies and societies, and the International Astronomical Union issued public statements[1]. The community quickly became aware of the planned developments of many other companies and nations to develop similar constellations.

A satellite constellation is defined as "a number of similar satellites, of a similar type and function, designed to be in similar, complementary, orbits for a shared purpose, under shared control"[2]. Currently operating constellations serve a variety of important and critical functions for society including: navigation and geodesy (e.g. GPS, Galileo and GLONASS), satellite telephony (e.g. Iridium), internet and TV (e.g. ViaSat, Orbcom, GlobalStar) and Earth Observation (e.g. Copernicus, DMC, PlanetLabs). In the future, companies such as SpaceX, Amazon, Samsung, Telesat, OneWeb and several national entities (Chinese and Indian Space Agencies) are planning very large constellations in Low Earth Orbit (LEO) to provide low-latency consumer broad-band internet around the world, support 'Internet of Things' to connect machines and systems together directly, financial and gaming transactions, and military applications[3].

---

[1] See, for example, ESO's official statement. A complete list is in the Reference Section.

[2] Wood, Lloyd, Satellite constellation networks, Internetworking and Computing over Satellite Networks. Springer, Boston, MA, 2003, p.13-34.

[3] Curzi, G., Modenini, D., & Tortora, P. (2020). Large Constellations of Small Satellites: A Survey of Near Future Challenges and Missions. *Aerospace*, 7(*9*), 133. https://doi.org/10.3390/aerospace7090133

| Constellation (Registering Nation) | Altitude (km) | Number of satellites |
|---|---|---|
| Starlink Generation 1 (US) | | |
| | 5,50 | 1,584 |
| | 1,110 | 1,600 |
| | 1,130 | 400 |
| | 1,275 | 375 |
| | 1,325 | 450 |
| Starlink Generation 2 (US) | | |
| | 328 | 7,178 |
| | 334 | 7,178 |
| | 345 | 7,178 |
| | 373 | 1,998 |
| | 499 | 4,000 |
| | 604 | 144 |
| | 614 | 324 |
| | 360 | 2,000 |
| OneWeb Phase 2 | | |
| (US, UK) | 1,200 | 1,764 |
| | 1,200 | 23,040 |
| | 1,200 | 23,040 |
| Amazon Kuiper (US) | | |
| | 590 | 784 |
| | 610 | 1,296 |
| | 630 | 1,156 |
| Sat Revolution (Poland) | 350 | 1,024 |
| CASC Hongyan (China) | 1,100 | 320 |
| CASIC Xingyun Lucky Star (China) | 1,000 | 156 |
| CommSat (China) | 600 | 800 |
| Xinwei (China) | 600 | 32 |
| AstromeTech (India) | 1,400 | 600 |
| Boeing (US) | 1,030 | 2,956 |
| LeoSat (Luxembourg) | 1,423 | 108 |
| Samsung (Korea) | 2,000 | 4,700 |
| Yaliny (Russia) | 600 | 135 |
| Telesat LEO (Canada) | 1,000 | 117 |
| Total | | 96,437 |

**Table 1: Planned Satellite Constellations.** *Starlink and OneWeb constellations include their updated architectures (both via May 2020 US Federal Communication Commission (FCC) filings). Starlink GEN1 reflects the currently licensed status. SpaceX has filed with the FCC a license modification request to bring the GEN1 satellites to orbits at altitudes between 540 and 570 km. At the time of preparing this document the FCC had not made a decision on accepting or rejecting the request. Some of the operators have withdrawn their applications (Boeing, LeoSat). Only Starlink and OneWeb are already assembling and launching satellites (over 830 and 74, respectively, as of Oct. 2020. Many more companies have filed applications for other purposes (e.g. remote sensing). As these are typically much smaller (hence fainter) than telecommunication satellites, they are not considered here*.

Drawing from public filings to the International Telecommunications Union (ITU) and also national regulatory agencies such as the United States' Federal Communications Commission (FCC), the number of planned satellite constellations and their numbers of individual units can be estimated. The numbers are estimates, as not all companies will likely succeed with their deployments, and in some cases ITU filings are aspirational or strategic rather than representing concrete projects.[4] From the planned LEO constellations numbering more than 500 units, only SpaceX and OneWeb have actually manufactured and launched satellites[5] and are by far the largest projects in terms of planned unit numbers. The following work considers only Starlink and OneWeb, as representative cases.

## 2. Impacts on ESO facilities and astronomical science in the Visible and IR

### 2.1. Method

Using the constellation parameters, one can compute representative positions of each satellite at a given time with a precision sufficient to evaluate their effect on observations. A faster alternative to this numerical brute force is to use analytical equations giving directly the density of satellites and related parameters for any azimuth and elevation. Using these methods, one can evaluate the number of satellite trails that will cross a field of view, as a function of the instrument characteristics, the pointing azimuth and altitude, the local time of the night and of the year. Figure 1 displays the result of one of these simulations. The methods are described in detail in the technical appendix of the SATCON1 report[6].

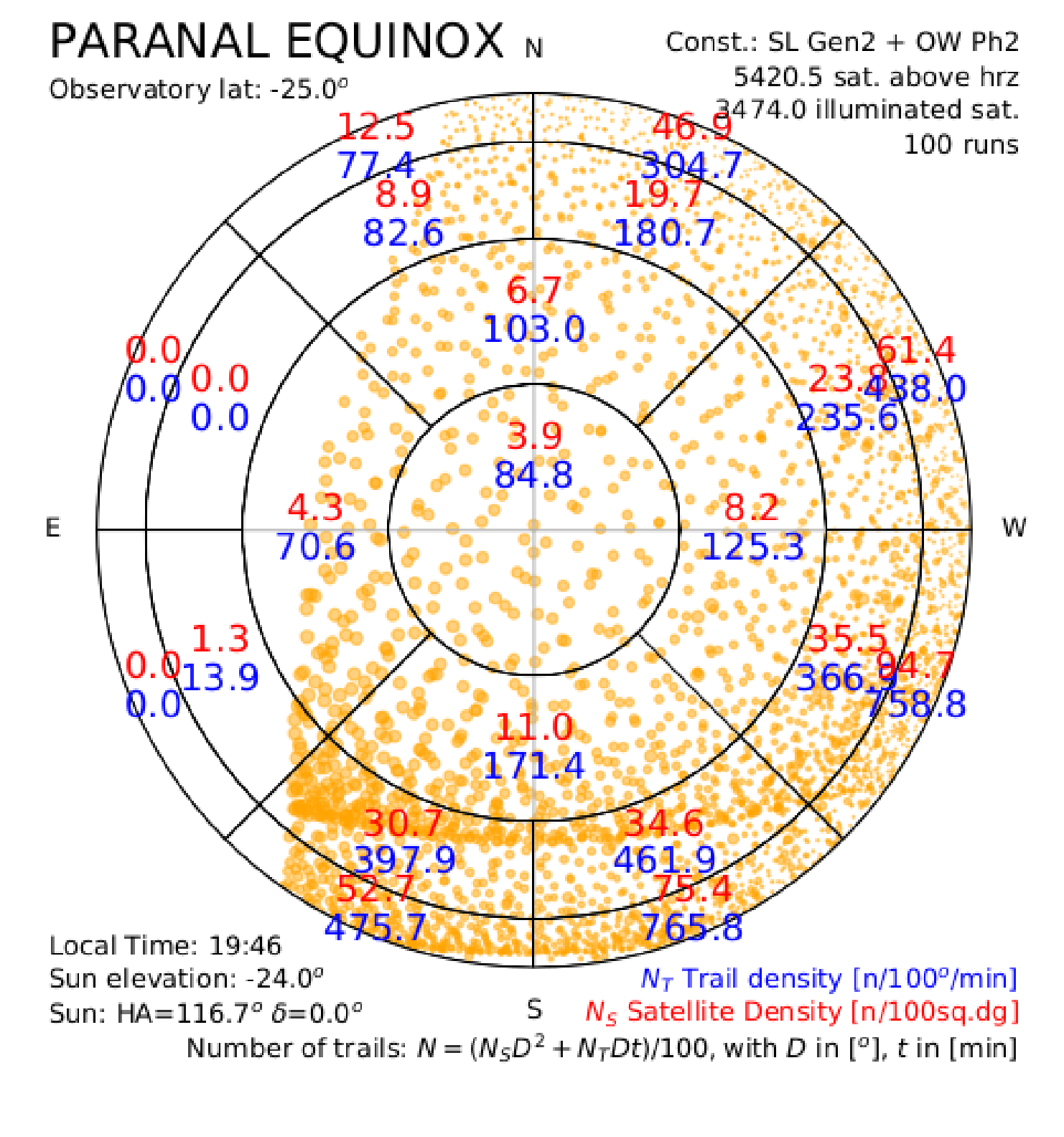

**Figure 1: Example distribution of satellites over Paranal**

*This figure shows a map of the sky above Paranal at the beginning of an equinox night, showing a representative distribution of Starlink (SL) and OneWeb (OW) satellites. Of the total 80,000 satellites, 5,420 are above the horizon, including 2,474 illuminated satellites. The numbers in the map give the density of satellites (in 100/sq.deg) and of trails (in 100/deg/min) in various regions of the sky. Such simulations were used to compute the effect of the SL+OW constellation discussed in this report.*

---

[4] https://spacenews.com/itu-wants-megaconstellations-to-meet-tougher-launch-milestones/ ; https://www.spacelegalissues.com/orbital-slots-and-space-congestion/

[5] Several other companies have launched individual prototypes, e.g. https://spacenews.com/telesat-preparing-for-mid-2020-constellation-manufacturer-selection/

[6] Walker, C. & Hall, J. (Eds.) (2020).

The constellation used for the simulations presented below is composed of Starlink Gen. 2 and OneWeb Phase 2 (SL+OW), totalling almost 80,000 satellites over a range of low and high orbits. This combination is representative of the overall population of satellites expected to be launched over the next decade. The effect on observations scales with the number of satellites, so the effect of the current ~800 constellation satellites is about 1% of what is discussed below.

The numbers reported below are the average for observations performed at elevations above 30 degrees (airmass 2 and better, representing 95% of VLT observations). As a result of the apparent concentration of satellites at low elevation, the numbers are about double those in the table for elevations above 20 degrees, and about half above 60 degrees.

The number of illuminated satellites depends strongly on the elevation of the Sun. The table below shows the value as a function of that elevation, as well as the average over the astronomical night. Note that the first and last hour of the night dominate in this average.

The photometric model used to evaluate the magnitude of the satellites is based on a simple geometric model calibrated on the Starlink Generation 1 satellites[7]. It must be noted that the current Generation 2 is about 1 magnitude fainter[8], and that the OneWeb satellites are smaller and fainter. *The model is therefore overestimating the brightness of the real satellites—or overestimating the number of bright satellites*. Furthermore, the fact that the satellites appear both resolved and defocused is not accounted for: the actual surface brightness of the satellites is therefore lower than that of point sources of the same magnitude[9]. As more satellite photometric measurement become available in the coming months and years, the photometric model will be refined, and will be adapted to the various types of satellites[10].

In order to evaluate the effect of satellite trails on observations, a few assumptions must be made. While the field-of-view of the instrument is defined, the exposure times vary widely depending on the science case: a typical exposure time is therefore chosen.

For several instruments, two cases are distinguished according to the brightness of the satellite: that of a bright object (mag V < 5, using the photometric model described above), and any satellite (whose magnitude is in the V= 3-14 range, with a peak at V ~ 9). Because of the very fast apparent motion of the satellite (10-70 deg/min, the faster being the brighter), the effective magnitude of the satellite (that is, the magnitude of a star that would have the same peak surface brightness) is 5-15 mag fainter. Also, low-altitude satellites are resolved, and appear out-of-focus in large telescope, further reducing their effective magnitudes.

It is assumed that a bright satellite crossing the field of view will ruin enough of the affected frame to consider it unusable. Fainter satellites will leave a trail across the field-of-view. The exposure level of that trail varies according to the spectral range, exposure time, observation

---

[7] Hainaut, O. R., & Williams, A. P. (2020).

[8] Tregloan-Reed, J. et al. (2020a); Tregloan-Reed, J. et al. (2020b); Tyson, J. A., et al., (2020).

[9] Ragazzoni, R. (2020).

[10] For example, recent observations show that the apparent magnitude of Starlink satellites increases from the optical to the near infrared spectral bands (Tregolan & Reed, 2020a)

technique, from bright to barely detectable. That trail potentially ruins a band a few arcsec wide (set to 5” for the discussion below). The effect on the usefulness of the affected frame depends on the observing technique and science case. In what follows, a frame with a bright satellite is considered “ruined”, while one with a faint satellite is marked as affected. Specific cases are discussed. Because of their very small number, bright satellites are not subtracted from the total of the faint ones.

## 2.2. ESO Facilities

**Overall effects**

The number of satellite trails decreases with the solar elevation below the horizon. Consequently, the number of affected frames is highest at twilight, and dwindles towards midnight. That effect is stronger and faster for bright satellites (which are at lower altitude, therefore falling faster in the shadow of the Earth). At twilight, the number of bright satellites is ~2.5% of the total number of satellites. That fraction drops to 0 when the Sun is below –24deg elevation.

Simply because of the geometry of the constellations, the number of satellite trails is much higher for fields at lower elevation above the horizon. As seen in Figure 1, there are about 10x more satellites between 10 and 30 deg elevation than above 60 deg.

**Classical imaging, represented by the case of FORS2**

Overall, 0.3% of the FORS2 images would be ruined by bright satellites (corresponding to 2.7% during astronomical twilight). Up to an average of 20% of the images would include a fainter satellite trail, each affecting up to 1.5% of the pixels of the affected image.

Beyond very short exposures, the effect scales as the exposure time and the diameter/linear size of the field of view. Instruments falling in this category include EFOSC2 (La Silla), FORS2, SPHERE, HAWK-I (Paranal).

**Wide-field imaging, represented by the case of OmegaCam on VST**

While OmegaCam will likely reach its end of life before the number of satellites becomes an issue, it is used to represent wide-field imaging. Overall, 2.6% of the frames would have a bright satellite trail across (corresponding to 27% at twilight), each potentially ruining up to 8 of the 32 detectors. On average, each 5min image would include 2 faint satellite trails, affecting 0.3% of the pixels (up to 7 satellite trails at twilight, affecting 1% of the pixels).

Beyond very short exposures, the effect scales as the exposure time and the diameter/linear size of the field of view. Instruments falling in this category include WFI (La Silla), OmegaCam, VIRCAM (Paranal). See also the special case of VRO/LSST below.

**Spectroscopy**

Because of the fast, apparent motion of the satellites, the object will remain in the slit (or fibre) only 0.2 to 2 milli-seconds. The exposure level of the resulting spectrum in a 5-min exposure will be around that of a 20-21 mag star (as the brighter satellites tend to move faster, the dependency on magnitude is weak); for a 20-min, the contamination is equivalent to a 22-23 mag star. The risk for spectroscopy is therefore to have spurious contamination rather than a frame blinded by a bright object.

For fibre-fed spectrograph, about 0.4% of the 20-min exposures would have a contamination (1.5% at twilight, of which 0.1% by a bright satellite). In the case of the multi-fibre 4MOST, simulations of the sparse coverage of the field-of-view indicate that each satellite trail crossing the field affects on average 1.25 fibres. For these spectrographs, which have no spatial information, the main risk is that the contamination might be un-noticed until the data are analysed. The issue of spectral contamination is being studied in detail by the 4MOST consortium[11].

In the case of fibre-fed spectrograph, the effect scales with the exposure time only. Instruments in this category include FEROS, HARPS, NIRPS (La Silla), FLAMES, KMOS, ESPRESSO, 4MOST, MOONS (Paranal).

For slit spectrographs, the number of satellite trails crossing the fibre scales with the length of the slit. For FORS, on average 60% of the 20-min exposures would have a satellite trail (up to 2 trails per exposure at twilight). In the case of UVES (with a much shorter slit), on average 2% of the 20-min exposures will have a satellite trail (up to 7% at twilight). Overall, the fraction of a long slit contaminated is constant: 1% of the slit on average, up to 3% at twilight. The effect on the science will depend on the science case.

In case of slit spectrograph, the effect scales with the length of the slit and the exposure time. Instruments in this category are EFOSC2, SOXS (La Silla), FORS2, XSHOOTER.

Integral Field Spectrographs are affected as imagers in terms of the number of satellite trails, and as spectrographs in terms of the effect of the contamination, with the advantage that the spatial information is preserved. Instruments in this category include FLAMES, MUSE and KMOS (Paranal).

**Thermal InfraRed (IR), represented by the case of VISIR**

In the thermal IR regime (5-20 micron range), the satellites emit considerable amounts of radiation. The satellite falling in the Earth's shadow don't change much that flux. Consequently, the values reported in Table 2 are constant over the various solar elevations.

Thermal IR imaging: A crude thermal model indicates that a satellite at 2000 km altitude would emit ~100 Jy in N and ~50 Jy in M and Q bands, which is easily detected on a single exposure with VISIR[12]. Individual exposures in the thermal IR are so short that the number of satellite scales with the area of the field of view and does not depend on the exposure time. In the case of VISIR, this results in 0.005% of the frames being affected. Accounting for the integrated exposure time (NDIT x DIT[13]), 0.2% of the images would contain a satellite.

---

[11] Micheva et al. (2020).

[12] Hainaut O. & Williams A.P. (2020).

[13] DIT is the exposure time of an individual exposure (very short in the Thermal IR, <<s), NDIT is the number of exposures acquired in a sequence to reach the total exposure time NDIT x DIT (typically seconds to minutes).

Because many images are acquired and median-combined, the effect in the Thermal IR imaging is therefore negligible.

Thermal IR spectroscopy: while the individual exposures are longer, resulting in more satellite trails, the observing technique also implies that many individual spectra are combined, filtering out the contamination by the satellites.

**VLTI**

The field-of-view of the VLTI instrument is tiny, resulting in a negligible number of contaminating satellites.

**Occultation**

Because of their very fast apparent motion, the occultation of an astronomical source by a satellite will last a very short time ($t_{occ}$ = 0.5 to 5ms depending on the altitude). The photometric effect will therefore be a reduction of the flux of $t_{occ}/t_{exp}$ ~ 0.5 to 5 millimag for a $t_{exp}$ = 1 s exposure. So, except for the fastest of fast photometry (currently not supported at ESO), the effect of occultation is completely negligible. Furthermore, the probability of this to happen is small: on average 0.3% of 10min sequences would include a satellite occultation.

**ELT instruments**

All first generation ELT instruments will have extremely small fields of view, compared to the VLT instruments. Consequently, the number of satellites trailing through the ELT instrument field of view will be much smaller than for the VLT. Furthermore, the larger size of the ELT will cause the image of the satellites to be even more resolved and out of focus than for the VLT[14]. Second generation instruments were not considered, but the same level of contamination is expected as the multi-object spectrographs for the VLT.

**Cherenkov Array Telescope**

The CTA telescopes have large field of view (4.3-9$^{o}$ diam.) paved with very large pixels (0.1-0.2$^{o}$). The individual exposure times are extremely short (ns), taken in a sequence of the order of 20 min. The observations can take place only in dark conditions. A detection can be triggered when a source brighter than a threshold is identified in one of the individual exposures, and confirmed in subsequent exposures, accounting for the geometry of a Cherenkov event. Stars brighter than the threshold are flagged, as their motion through the field-of-view corresponds to the parallactic field rotation during the sequence.

Satellites brighter that the detection threshold will move in a different fashion, crossing the field of view in few seconds to 1 min. They will have to be identified and flagged so not to be counted as Cherenkov events. Satellites fainter than the threshold cannot trigger a detection but contribute to the background illumination of the pixels.

For observations above airmass 2, an average of 30 satellite trails will cross the CTA LST FoV during a 20 min sequence (70 for SST). These numbers are dominated by the crossing taking place during the first or last half-hour of the night. The average spatial density of satellites is about the same as that of galactic stars around the galactic pole for objects brighter than 8, and much lower for fainter objects.

---

[14] Ragazzoni, R. (2020).

The threshold magnitude is ~5 or brighter. In what follows, the bright satellites limit is set to V=6. An average of 5 (LST) or 10 (SST) bright satellites can be expected to cross the field of view of the telescope during the first and last half-hour of the nigh. In between, no bright satellites are expected, at all.

A more detailed study of the effect of satellites on CTA telescopes is being performed by the CTA consortium.

### 2.3. Impacts on Other Observatories

The effect on other facilities can be extrapolated from Table 2, comparing with similar ESO instruments. Some special cases deserve to be highlighted.

**Special Case: Vera Rubin Observatory**

The Legacy Survey of Space and Time (LSST) performed on the large (8 m) Vera C. Rubin Observatory telescope with its huge CCD mosaic is a special case: because of the large field of view and 5min exposures, a large number of satellite trails is expected to streak the exposures (an average of 5.7 streaks, but with a peak above 20 trails per image at twilight).

In addition to the loss of efficiency for the survey caused by the contaminated pixels, the camera suffers from electronic and optical ghosts when a satellite brighter than V~7 crosses the field of view, rendering useless the data from the whole affected quadrant of the camera. While bright satellites are an issue only during the beginning or end of night, a large fraction of the frames taken during that period could be affected. See the mitigation section.

**Low elevation surveys, twilight surveys**

Because of the increased number of satellites during twilight (1 hr after twilight there are 2x less satellites, and virtually no bright satellites), observations that must be performed during twilight will be more severely affected. Furthermore, as the shadow of the Earth is in the direction opposite to the Sun, observations that need to be performed during twilight in the direction of the Sun will be even more severely affected. Finally, observations performed at low elevation always see a higher apparent density of satellites.

Therefore, the numbers for wide-field surveys observing at low elevation and low solar elongation could be an order of magnitude higher than what is reported in this document. This would be the case for some survey for comets and NEOs.

### 2.4. Mitigation Strategies

Various levels of mitigations can be considered.

**Actions on the satellites**

*Satellites number and altitude.*

These obvious parameters are mentioned for completeness: fewer satellites would decrease the effect, and satellites on lower orbit make their effect confined to the twilight period. However, the architecture of a constellation (number of satellites, number and altitude and inclination of the sub-constellations) is finalised very early in the definition of a constellation, and it is unlikely that a satellite operator would (or even could) modify them.

*Brightness of the satellites*

Decreasing the brightness of a satellite can be achieved in various ways: making them smaller is in theory an option, but unpractical. Making them darker is a possibility. Starlink's DarkSat experiment and the measurements of that satellite indicate that –all other things being equal- DarkSat is about 2x fainter in the visible than the other satellites[15]. This factor however is less good at longer wavelengths[16]. Also, the darkening of the satellites has implications on their thermal control.

The satellite can also be made fainter by adjusting its attitude. This is done since a few months by Starlink for the satellites on their transit orbit, after launch until they reach their operational altitude. By rotating the satellite so that its narrow edge points towards the Sun, a significant reduction of brightness is achieved. Thanks to this, the very bright "string of pearls" that were the hallmark of early Starlink launches have not been reported for the recent launches. Once on their operational orbit, the position of the solar panel can be adjusted so that the illuminated side is not visible from Earth (done by Starlink).

Finally, the satellite can be equipped with a sunshade which protects the body of the satellite from direct illumination, and which is mounted so that the illuminated side of the shade is not visible from Earth. This is the principle of the Starlink "VisorSat," which is being launched since mid-2020. Preliminary measurements suggest a darkening by a factor 2-3 compared to first generation satellites[17]. Starlink is experimenting with new, improved designs of the sunshade.

Starlink committed to reducing the brightness of their satellites so they are fainter than V~7[18], this limit corresponding to the saturation threshold of the Vera C. Rubin Observatory's telescope camera.

**Actions on the telescopes, instruments, and their operation:**

*Statistical scheduling*

Satellite density maps such as the one displayed in Figure 1 can be used to fine-tune the short-term scheduling of the observations, in particular in Service Mode. By pointing in the regions of the sky devoid of satellites (because they are in the shadow of the Earth), or at least avoiding the regions of the sky with the highest density, the number of trails can be significantly reduced. The density maps can easily be generated in real-time (thanks to the analytical modelling of the constellations developed by Bassa et al. (2020)) and could be integrated in the telescope short-term scheduler. While less exposures would be affected by satellite trails, the efficiency of the scheduling could be degraded (e.g. resulting in longer slew time). The gain of unaffected exposure has to be balanced with that loss of efficiency.

[15] Tregloan-Reed, J. et al. (2020a).

[16] Tregloan-Reed, J. et al. (2020b).

[17] Otarola, A. et al. (2020).

[18] Walker, C. & Hall, J. (Eds.) (2020)

*Deterministic scheduling*

For critical observation, the exact timing and position of the interfering satellites can be computed in advance, and this information used to reschedule an observation.

This would require having all the orbital elements of all satellites and computing their position in real time. While some satellite operators committed to making the orbital elements available with a sufficient precision, it is unclear whether that information would be available for all satellites (the traditional "two-line elements" widely available for all satellites are not precise enough).

*Shutter control*

An alternative to the previous mitigation method would be to close the shutter just before a satellite enters the field of view, and re-open it when it left. This could be done computing the position of all satellites, with the same caveat as above. A more promising alternative would be to equip each telescope with a small camera mounted in parallel, to monitor the field of view, detect incoming satellites, and trigger the shutter. Such a monitoring system would also be valuable for wide-field multi-fibre spectrographs like 4MOST. It would keep a record of which fibres are contaminated by a satellite trail. This method, like the aircraft detection camera used for the Laser guide stars, has the advantage not to rely on external information.

The level and complexity of telescope or operations mitigation put in place must be commensurate to the effect it corrects. For many types of observations, it might be cheaper and more efficient just to repeat a failed observation. Currently, and until several thousands of satellites are launched, the impact on ESO facilities will remain low enough that no telescope/instrument nor scheduling mitigations are foreseen.

**Table 2: Summary of Effects on ESO and Other Facilities.**

Table 2 shows the number of satellite trails per exposures for a variety of representative exposures with diverse instruments, computed for the 80,000 low- and high-altitude of the SL+OW constellations, and for most instruments, the total number of bright satellites trails (mag < 6, likely to ruin all or a large fraction of the detector area for that exposure), and the total number of satellites (making a trail but affecting only the pixels below the trail). The sizes (width and length, or diameter) of the field of view, and typical exposure times are listed for each. For 4MOST, the number of trails in the field is converted into number of affected fibres using a Monte-Carlo simulation which indicates that on average a satellite will affect 1.254 fibre. All calculations were performed for Paranal (rounding to latitude 25° S)

The effect on the corresponding science exposures is colour-coded as follow. Green ●: less than 0.1%, yellow-grey ●: up to 1%, bright yellow ●: up to 3%, light red ●: up to 30%, dark red ●: above 30% The number of trails is given as a function of the solar elevation below the horizon; twilights are marked in colour. The corresponding local solar time is given for the solstices and equinoxes. The average number of trails over the astronomical night at equinox is listed in the shaded column. Twilight exposures that are unlikely to be performed are greyed out. For this table, exposures above an elevation of 30deg are considered. Lowering the limit to 20deg about doubles the values, raising it to 60deg about halves the values.

| | length | width | ExpTime | Trails per exposure | Sunset | Civil | Nautical | Astron. | Night... | | | | | | | | |
|---|---|---|---|---|---|---|---|---|---|---|---|---|---|---|---|---|---|
| Sun Elevation [deg]: | | | | | 0 | -6 | -12 | -18 | -24 | -30 | -36 | -42 | -48 | -54 | -60 | -66 ... | -84 |
| Local Solar Time [hh:mm]: Winter | | | | | 17:14 | 17:43 | 18:11 | **18:39** | 19:07 | 19:34 | 20:01 | 20:28 | 20:55 | 21:22 | 21:48 | 22:15 ... | **00:00** |
| Equinox | | | | | 18:00 | 18:26 | 18:53 | **19:19** | 19:46 | 20:13 | 20:41 | 21:10 | 21:40 | 22:13 | 22:51 | **00:00** | ////////////// |
| Summer | | | | | 18:45 | 19:15 | 19:46 | **20:18** | 20:53 | 21:31 | 22:18 | **00:00** | ////////////// | ////////////// | ////////////// | ////////////// | ////////////// |
| Field of view | length | width | ExpTime | **Trails per exposure** | | | | | | | | | | | | | |
| **VLT** | | | | Average over equinox night | | | | | | | | | | | | | |
| **FORS Imaging** | | | | | | | | | | | | | | | | | |
| All sat. | 6' | 6' | 5min | 0.19 | 0.736 | 0.746 | 0.746 | 0.681 | 0.546 | 0.400 | 0.212 | 0.052 | 1.4E-04 | 1.8E-06 | 0 | 0 | 0 |
| Bright sat. | | | | 0.0026 | 0.078 | 0.081 | 0.082 | 0.027 | 0 | 0 | 0 | 0 | 0 | 0 | 0 | 0 | 0 |
| **FORS Spectroscopy** | | | | | | | | | | | | | | | | | |
| All sat. | 5' | 2" | 20min | 0.62 | 2.450 | 2.483 | 2.483 | 2.267 | 1.817 | 1.332 | 0.707 | 0.173 | 4.8E-04 | 6.1E-06 | 0 | 0 | 0 |
| Bright sat. | | | | 0.009 | 0.260 | 0.268 | 0.273 | 0.089 | 0 | 0 | 0 | 0 | 0 | 0 | 0 | 0 | 0 |
| **UVES high-res. Spectro.** | | | | | | | | | | | | | | | | | |
| All sat. | 10" | 2" | 20min | 0.021 | 0.082 | 0.083 | 0.083 | 0.076 | 0.061 | 0.044 | 0.024 | 0.006 | 1.6E-05 | 2.0E-07 | 0 | 0 | 0 |
| Bright sat. | | | | 0.00028 | 0.009 | 0.009 | 0.009 | 0.003 | 0 | 0 | 0 | 0 | 0 | 0 | 0 | 0 | 0 |
| **ESPRESSO HARPS fibre-fed spectro.** | | | | | | | | | | | | | | | | | |
| All sat. | 2" | | 20min | 0.0041 | 0.016 | 0.017 | 0.017 | 0.015 | 0.012 | 0.009 | 0.005 | 0.001 | 3.2E-06 | 4.1E-08 | 0 | 0 | 0 |
| Bright sat. | | | | 0.000057 | 0.002 | 0.002 | 0.002 | 0.001 | 0 | 0 | 0 | 0 | 0 | 0 | 0 | 0 | 0 |
| **Occultation/Transit/VLTI** | | | | | | | | | | | | | | | | | |
| All sat. | 3" | | 10min | 0.0031 | 0.012 | 0.012 | 0.012 | 0.011 | 0.009 | 0.007 | 0.004 | 0.001 | 2.4E-06 | 3.1E-08 | 0 | 0 | 0 |
| **VISIR Thermal IR** | | | | | | | | | | | | | | | | | |
| Single DIT | 60" | | ms | 0.000046 | 4.6E-05 | 4.6E-05 | 4.6E-05 | 4.6E-05 | constant with solar elevation | | | | | | | | |
| Frame | 60" | | 5s | 0.0021 | 2.1E-03 | 2.1E-03 | 2.1E-03 | 2.1E-03 | constant with solar elevation | | | | | | | | |
| **Wide-Field** | | | | | | | | | | | | | | | | | |
| **VST Wide-field imaging** | | | | | | | | | | | | | | | | | |
| All sat. | 1deg | | 5min | 1.9 | 7.429 | 7.530 | 7.530 | 6.877 | 5.516 | 4.043 | 2.148 | 0.528 | 1.5E-03 | 1.8E-05 | 0 | 0 | 0 |
| Bright sat. | | | | 0.026 | 0.783 | 0.808 | 0.823 | 0.267 | 0.001 | 0 | 0 | 0 | 0 | 0 | 0 | 0 | 0 |
| **4MOST multi-fibre spectro.** | | | | | | | | | | | | | | | | | |
| All satellites | | | | | | | | | | | | | | | | | |
| Full FoV | 2.3deg | | 20min | 17.1 | 68.037 | 68.962 | 68.962 | 62.967 | 50.487 | 37.010 | 19.651 | 4.826 | 1.4E-02 | 1.7E-04 | 0 | 0 | 0 |
| Number of fibers affected | | | | 21.5 | 85.319 | 86.478 | 86.479 | 78.960 | 63.311 | 46.411 | 24.642 | 6.052 | 0.017 | 0 | 0 | 0 | 0 |
| % of fibers | | | | 0.0088 | 3.50% | 3.55% | 3.55% | 3.24% | 2.60% | 1.91% | 1.01% | 0.25% | 0.00% | 0% | 0% | 0% | 0% |
| Bright satellites | | | | | | | | | | | | | | | | | |
| Full FoV | 2.3deg | | 20min | 0.24 | 7.193 | 7.423 | 7.562 | 2.448 | 0.007 | 0 | 0 | 0 | 0 | 0 | 0 | 0 | 0 |
| Number of fibers affected | | | | 0.30 | 9.020 | 9.309 | 9.483 | 3.070 | 0.009 | 0 | 0 | 0 | 0 | 0 | 0 | 0 | 0 |
| % of fibers | | | | 0.00012 | 0.37% | 0.38% | 0.39% | 0.13% | 0.00% | 0% | 0% | 0% | 0% | 0% | 0% | 0% | 0% |

| | Field of view length | width | ExpTime | Trails per exposure | Sunset | Civil | Nautical | Astron. | Night… | | | | | | | | |
|---|---|---|---|---|---|---|---|---|---|---|---|---|---|---|---|---|---|
| Sun Elevation [deg]: | | | | | 0 | -6 | -12 | -18 | -24 | -30 | -36 | -42 | -48 | -54 | -60 | -66 … | -84 |
| Local Solar Time [hh:mm]: | | | | Winter | 17:14 | 17:43 | 18:11 | **18:39** | 19:07 | 19:34 | 20:01 | 20:28 | 20:55 | 21:22 | 21:48 | 22:15 … | **00:00** |
| | | | | Equinox | 18:00 | 18:26 | 18:53 | **19:19** | 19:46 | 20:13 | 20:41 | 21:10 | 21:40 | 22:13 | 22:51 | **00:00** ////////////▸ | |
| | | | | Summer | 18:45 | 19:15 | 19:46 | **20:18** | 20:53 | 21:31 | 22:18 | **00:00** ////////////////////////////////////////////////////////▸ | | | | | |
| **ELT** | | | | Average over equinox night | | | | | | | | | | | | | |
| **HARMONI IFU** | | | | | | | | | | | | | | | | | |
| All sat. | 9.1" | 6.2" | 10min | 0.0092 | 0.037 | 0.037 | 0.037 | 0.0340 | 0.0273 | 0.0200 | 0.0106 | 0.0026 | 7.2E-06 | 9.2E-08 | 0 | 0 | 0 |
| Bright sat. | | | | 0.000128 | 0.0039 | 0.0040 | 0.0041 | 0.0013 | 0 | 0 | 0 | 0 | 0 | 0 | 0 | 0 | 0 |
| **METIS** | | | | | | | | | | | | | | | | | |
| **Thermal IR Imaging** | | | | | | | | | | | | | | | | | |
| All sat. | 10" | | 1min | 0.0041 | 0.004 | 0.004 | 0.004 | 0.004 | constant with solar elevation | | | | | | | | |
| Bright sat. | | | | 0.00045 | 0.0004 | 0.0004 | 0.0004 | 0.0004 | constant with solar elevation | | | | | | | | |
| **Thermal IR long-slit spectro.** | | | | | | | | | | | | | | | | | |
| All sat. | 10" | 0.1" | 5min | 0.02069 | 0.02069 | 0.02069 | 0.02069 | 0.02069 | constant with solar elevation | | | | | | | | |
| Bright sat. | | | | 0.002236 | 0.00228 | 0.00224 | 0.00228 | 0.00228 | constant with solar elevation | | | | | | | | |
| **Thermal IR IFU spectro.** | | | | | | | | | | | | | | | | | |
| All sat. | 0.25" | | 5min | 0.00046 | 0.00046 | 0.00046 | 0.00046 | 0.00046 | constant with solar elevation | | | | | | | | |
| Bright sat. | | | | 0.000049 | 0.00005 | 0.00005 | 0.00005 | 0.00005 | constant with solar elevation | | | | | | | | |
| **MICADO** | | | | | | | | | | | | | | | | | |
| **Imaging** | | | | | | | | | | | | | | | | | |
| All sat. | 50" | | 5min | 0.0257 | 0.102 | 0.103 | 0.103 | 0.0945 | 0.0757 | 0.0555 | 0.0294 | 0.0072 | 2.0E-05 | 2.5E-07 | 0 | 0 | 0 |
| Bright sat. | | | | 0.00036 | 0.0108 | 0.0112 | 0.0114 | 0.0037 | 0 | 0 | 0 | 0 | 0 | 0 | 0 | 0 | 0 |
| **Long-Slit spectro.** | | | | | | | | | | | | | | | | | |
| All sat. | 3" | 0.1" | 10min | 0.0031 | 0.012 | 0.012 | 0.012 | 0.0113 | 0.0091 | 0.0067 | 0.0035 | 0.0009 | 2.4E-06 | 3.1E-08 | 0 | 0 | 0 |
| Bright sat. | | | | 0.00004 | 0.0013 | 0.0013 | 0.0014 | 0.0004 | 0 | 0 | 0 | 0 | 0 | 0 | 0 | 0 | 0 |
| **Other facilities** | | | | | | | | | | | | | | | | | |
| **VRO/LSST Ultra-wide-field imaging** | | | | | | | | | | | | | | | | | |
| All sat. | 3deg | | 5min | 5.7 | 22.760 | 23.067 | 23.068 | 21.092 | 16.940 | 12.421 | 6.610 | 1.631 | 5.2E-03 | 5.5E-05 | 0 | 0 | 0 |
| Bright sat. | | | | 0.078 | 2.369 | 2.445 | 2.491 | 0.806 | 0.002 | 0 | 0 | 0 | 0 | 0 | 0 | 0 | 0 |
| **CTA:** | | | | | | | | | | | | | | | | | |
| **SST** | | | | | | | | | | | | | | | | | |
| **Full FoV – single exposure** | | | | | | | | | | | | | | | | | |
| All sat. | 4.3deg | | ns | 0.41 | 1.459 | 1.474 | 1.476 | 1.422 | 1.213 | 0.895 | 0.514 | 0.146 | 1.8E-03 | 0 | 0 | 0 | 0 |
| Bright sat. | | | | 0.0018 | 0.059 | 0.061 | 0.063 | 0.018 | 0 | 0 | 0 | 0 | 0 | 0 | 0 | 0 | 0 |
| **Full FoV – Long sequence** | | | | | | | | | | | | | | | | | |
| All sat. | 4.3deg | | 20min | 32.2 | 127.879 | 129.614 | 129.616 | 118.382 | 94.953 | 69.609 | 36.978 | 9.090 | 2.7E-02 | 0 | 0 | 0 | 0 |
| Bright sat. | | | | 0.44 | 13.475 | 13.907 | 14.167 | 4.585 | 0.013 | 0 | 0 | 0 | 0 | 0 | 0 | 0 | 0 |
| **SST** | | | | | | | | | | | | | | | | | |
| **Full FoV – single exposure** | | | | | | | | | | | | | | | | | |
| All sat. | 9deg | | ns | 1.80 | 6.391 | 6.456 | 6.464 | 6.229 | 5.314 | 3.920 | 2.252 | 0.640 | 8.1E-03 | 0 | 0 | 0 | 0 |
| Bright sat. | | | | 0.0078 | 0.259 | 0.267 | 0.275 | 0.081 | 0.000 | 0 | 0 | 0 | 0 | 0 | 0 | 0 | 0 |
| **Full FoV Long exposure** | | | | | | | | | | | | | | | | | |
| All sat. | 9deg | | 20min | 68.4 | 270.991 | 274.656 | 274.664 | 251.029 | 201.514 | 147.740 | 78.572 | 19.360 | 6.0E-02 | 6.6E-04 | 0 | 0 | 0 |
| Bright sat. | | | | 0.93 | 28.339 | 29.247 | 29.795 | 9.639 | 0.028 | 0 | 0 | 0 | 0 | 0 | 0 | 0 | 0 |

## 2.5. Impacts on the Visibility of the Dark Sky

Citizen visual observations (with the unaided eye or through a telescope) and astrophotography are also going to be affected by satellite constellations, however, not at the level that has been reported in some media—there will not be more satellites visible in the sky than bright stars.

Using the same simulations as for the telescopic observations, one can estimate the number of satellites and their impact. The number of satellites above the horizon is ~5% of their total number. Of these, about half are clustered at very low elevation (<10deg), where the horizon haze and obstructions make them less problematic. Of these satellites in sight, only those illuminated by the sun could be visible, and only those bright enough to be visible will contribute to the night sky pollution. Running the numbers for the 80,000 Starlink and OneWeb satellites indicate that ~2,500 satellites would be present in the sky over a mid-latitude place (Brussels, rounded to latitude 50 deg N), all of them illuminated at sunset (see Figure 2 for the plots). That number drops to half that value 1 hour after twilight and reaches zero 4 hours after twilight. At equinox, the night is not long enough, so that some satellites remain illuminated during the whole night. In summer, there is actually no astronomical night.

Of these satellites, only 35 will be brighter than mag = 5 at nautical twilight. One hour later, no satellite is left brighter than mag = 6. Consequently, as today, spotting bright satellites will remain fairly rare, with only a handful of objects visible during twilight (as a comparison, with the existing satellites in 2019, from 5 to 10 are reported visible and brighter than mag = 5).

In terms of wide-angle astrophotography (i.e., landscape photography, not through a telescope), the trailing of the object during the exposure reduces the effective limiting magnitude of the exposure for the satellites. For example, a 60 s exposure will only register satellites 7 to 10 mag brighter than the faintest star visible. Consequently, only astrophotographs capable of recording stars fainter than mag 12 in a 60 s exposure will be affected, and the number and period of visibility of the satellites are similar as for the unaided eye case.

Another source of concern are the satellites during the period extending from their launch until they reach their operational altitude. During that time, they are on extremely low orbits, and for Starlink and similar satellites, that period can extend for 4-8 weeks. Because of their lower altitude, the satellite will appear brighter. Furthermore, as many satellites are launched simultaneously, they will appear as a train, or “string of pearls”, for a few weeks until they spread over their orbit. The extremely bright (mag 0-2) early Starlink satellite trains were indeed spectacular and caused extreme concern among stargazers. Since then SpaceX has changed the attitude of the satellites on their transit orbit from one that maximized their cross section to another one that minimizes it, resulting in barely visible satellite trains. Extrapolating the number of launches required to assemble and then maintain the constellation, one can expect of the order of 1-4 compact train will be present around the Earth at any time. A given observer watching the sky every evening, would see one of these trains every few nights. The chances to catch one on camera without intending to are therefore small.

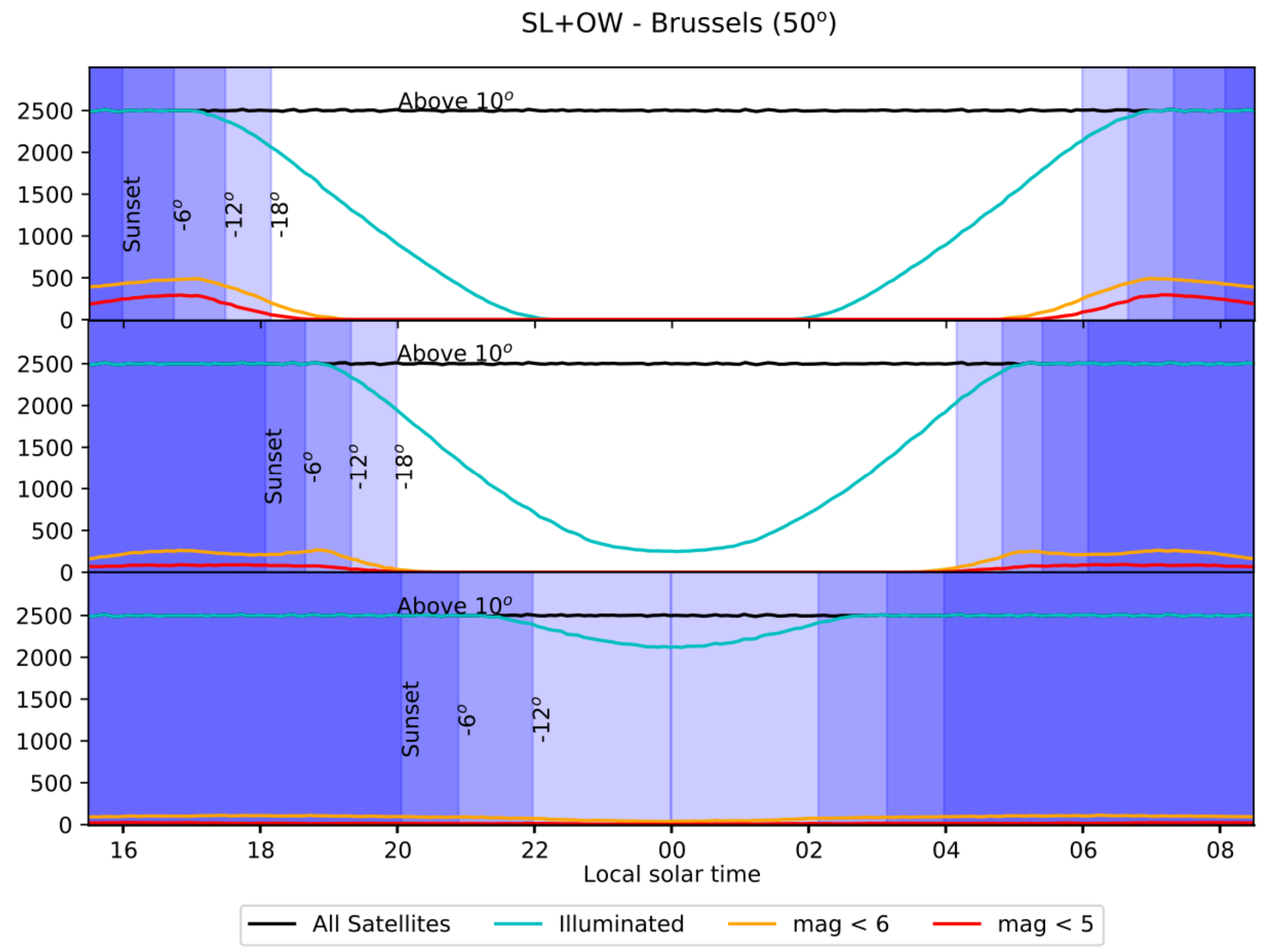

**Figure 2: Number of satellites above 10 degrees of elevation, as a function of the local time.** *The black line indicates all the satellites, the cyan one only those that are illuminated by the Sun, the red line those brighter than mag = 5 (visible in good conditions) and the orange one those brighter than mag = 6 (visible only in perfect conditions and completely dark skies). The blue shadings mark the twilights. These numbers are for 50$^{0}$ latitude*.

## 3. Impacts on ALMA

### 3.1. Introduction

Interference from satellite constellations has been a known issue for radio telescopes for more than 20 years. Already in the last decade of the previous century radio observatories encountered interference from the Iridium constellation in the 18 cm OH band. Due to the short observing wavelengths of ALMA, below 10 mm, it has been in the fortunate situation that so far almost no RFI issues have been encountered. However due to advancing technology developments and increasing societal needs for broadband internet access, also in remote areas, this will change in the coming decades. With the increased occupancy of the RF spectrum at longer wavelengths and the demand for high bandwidth for broadband internet, satellite constellations like Starlink are moving into the wavelength regime where ALMA is observing.

## 3.2. Regulatory framework

In contrast to the optical wavelength range, radio spectrum usage has been regulated at international level for more than 100 years. Since 1906 the International Telecommunication Union (ITU) is playing a pivotal role with establishing the Radio Regulations (RR) which governs the spectrum allocations for each service in the frequency range from 9 kHz to, currently, 3 THz. Every few years these RR are reviewed and updated if necessary by the ITU in consultation with the stakeholders, primarily national governments but also user representatives. It is noted that these representatives include satellite operators like Starlink but also the radio astronomy community. The most recent updated of the RR was made in 2019 at a World Radio Conference (WRC) held in Sharm al Sheikh, Egypt. Detailed information on the spectrum allocations to RAS can be found in the ITU published RR available on the Internet[19]. A more narrative description, but not fully up-to-date, can be found in the Handbook for Radio Astronomy[20] published by the Committee on Radio Astronomy Frequencies (CRAF).

For each part of the frequency spectrum the RR defines the services that are allowed to operate. Both satellite systems and radio astronomy have been recognized by the ITU as services in this respect. Beyond this frequency allocation in the RR the ITU also makes recommendations on e.g. the acceptable level of mutual interference under defined conditions, e.g. antenna patterns. A very important recommendation for the Radio Astronomy Services (RAS) is publication ITU-R RA.769-2 "Protection criteria used for radio astronomical measurements"[21]. This recommendation defines which maximum level of interference from other, active services, e.g. satellite transmitters, is allowed in those parts of the frequency spectrum allocated to RAS. Unfortunately, only small fractions of the spectrum have been allocated to the RAS in which ALMA is operating. Most of the frequency range covered by ALMA has to be shared with other, active (i.e. transmitting) services like satellite constellations.

For radio interference from satellite constellations encountered by radio observatories two RFI cases must be distinguished:

1. Transmissions from satellites at their allocated frequencies but also in use by a radio telescope, and,
2. Transmissions from satellites- in the frequency spectrum allocated to the RAS.

For the largest satellite constellations listed in Table 1 only the two Starlink networks, Generation 1 and Generation 2, will operate in their allocated frequency bands but also in use by the ALMA observatory (so case 1). The highest transmission frequency foreseen for OneWeb satellites is 19.3 GHz, well outside the ALMA observing range, and not considered a threat.

---

[19] https://www.itu.int/pub/R-REG-RR

[20] https://craf.eu/wp-content/uploads/2015/02/CRAFhandbook3.pdf

[21] https://www.itu.int/rec/R-REC-RA.769/en

Having been granted the necessary licenses by national authorities these services have the full legal right to operate at their allocated frequencies under specified conditions. Radio observatories like ALMA in principal have to accept the interference. Note that it is considered interference by a radio observatory but is a desired signal by the service itself. Only by collaboration between the primary stakeholders, radio observatory, satellite operator and possibly national regulatory authorities, other arrangements can be agreed to mitigate this interference.

RFI case 2 can exist for ALMA since some of the transmission frequencies of Starlink satellites are directly adjacent to the protected RAS frequency bands observed by ALMA. Due to unavoidable anomalies in the transmitters, e.g. due to non-linear behaviour of amplifiers, spurious emission spilling over in the RAS allocated bands will occur. For this RFI case the transmitting service has to fulfil the stringent requirements as defined in the aforementioned recommendation ITU-R RA.769-2.

### 3.3. Starlink Generation 1 RFI impact analysis

Of the satellite constellations listed in Table 1 the Starlink Gen1 is expected to have biggest impact on ALMA. The two main reasons for this are a) the downlink transmitters for the end user terminals operate within the ALMA observing range and b) the large number of satellites (~4400). Based on the rather limited, but reliable technical information published by SpaceX in their Starlink Gen1 FCC application an impact analysis has been made of which a summary is provided below.

**Starlink Gen 1 spectrum allocation and usage**

The frequency range of interest is the V-band (40 to 75 GHz following IEEE standards) and in particular the sub-ranges 37,5 – 42,5 GHz, 42,5 – 43,5 GHz, and 47,2 – 52,4 GHz.

Table 3 below summarizes the V-band frequency ranges in which Starlink Gen 1 intends to operate.

**Table 3: V-band frequency ranges used by the SpaceX system**

| Type of Link and Transmission Direction | Frequency Ranges |
|---|---|
| Downlink Channels<br><br>Satellite to User Terminal or<br>Satellite to Gateway | 37.5 – 42.5 GHz |
| Uplink Channels<br><br>User Terminal to Satellite or<br>Gateway to Satellite | 47.2 – 50.2 GHz<br><br>50.4 – 52.4 GHz |
| TT&C[22] Downlink<br><br>Beacon | 37.5 – 37.75 GHz |

[22] TT&C – Telemetry, Tracking, and Control

| | |
|---|---|
| TT&C[3] Uplink | 47.2 – 47.45 GHz |

Figure 3 provides a more detailed, graphical representation of the foreseen Starlink frequency usage.

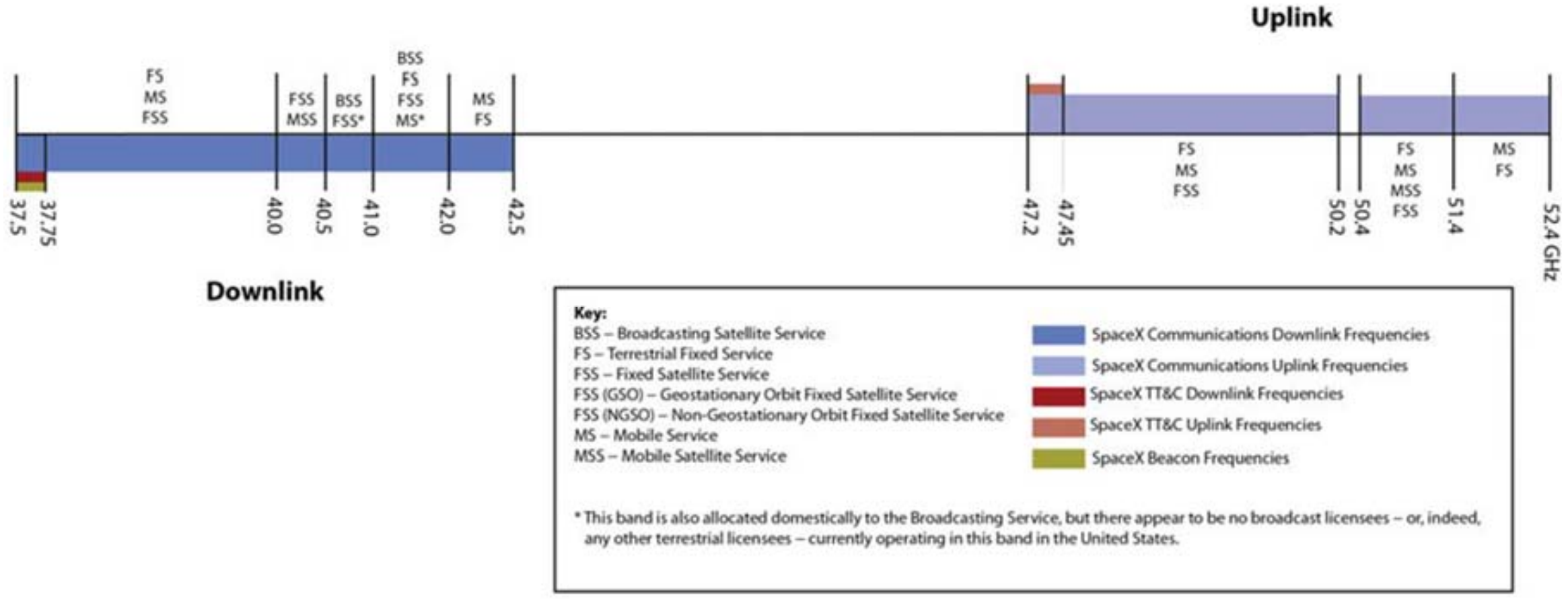

**Figure 3: V-Band spectrum used by the SpaceX system**

For the intersect between V-band (40 – 75 GHz) and ALMA Band 1 (30 – 50 GHz) the frequency range 42,5 to 43,5 GHz is of special interest since this is the only band where the ITU Radio Regulations define the Radio Astronomy Service (RAS) as primary user. In practice this means that other users have the obligation to minimize their, spurious, emissions to a very low level to allow the primary RAS user to operate.

### Starlink Gen 1 V-band downlink properties

For both User and Gateway downlinks the transmit beams are generated by phased array antennas mounted on the satellite. These phased array antennas generate spot beams which have a half-power beamwidth (HPBW), at nadir, of 1.0° for the Gateway link respectively 1.5° for the User link beams. These spot beams can be steered electronically by the phased array antenna. This beam steering is limited to about 51° from nadir for the VLEO satellites (Figure 5) and 44° for the LEO one's (Figure 4).

The satellite signal transmitted will have random, noise like properties unless a demodulation scheme matching the exact modulation parameters is applied on the receiving side. The latter is not applicable for ALMA and other radio telescopes.

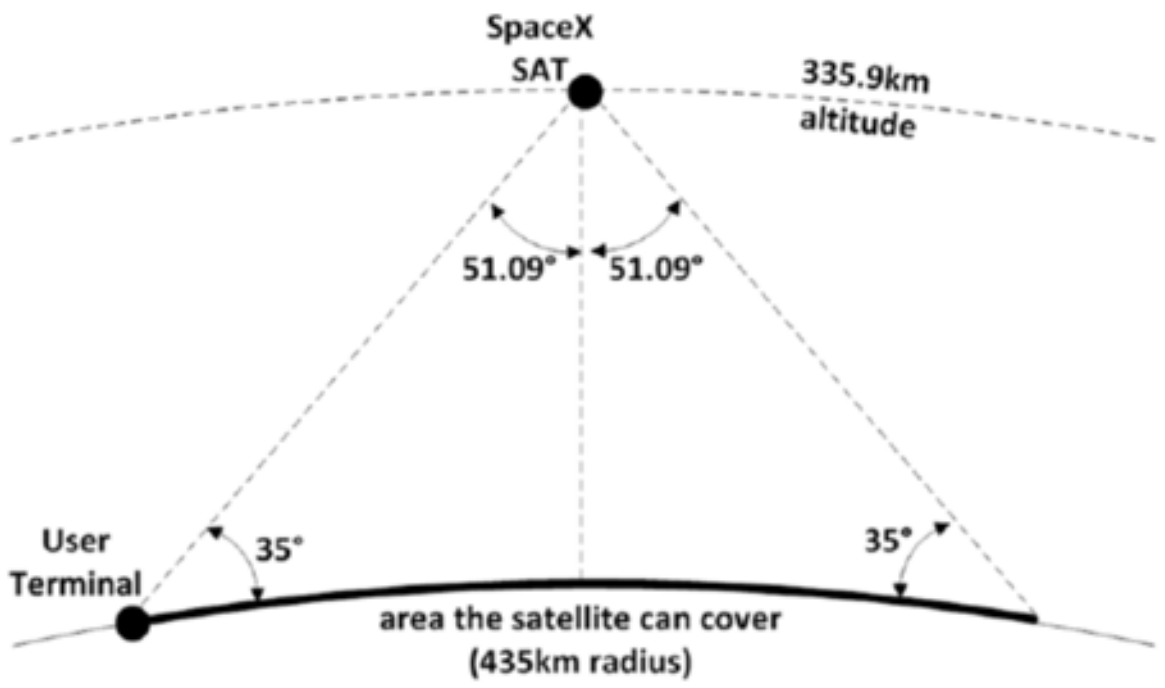

**Figure 5: Steerable Service Range of VLEO Beams (335.9 km)**

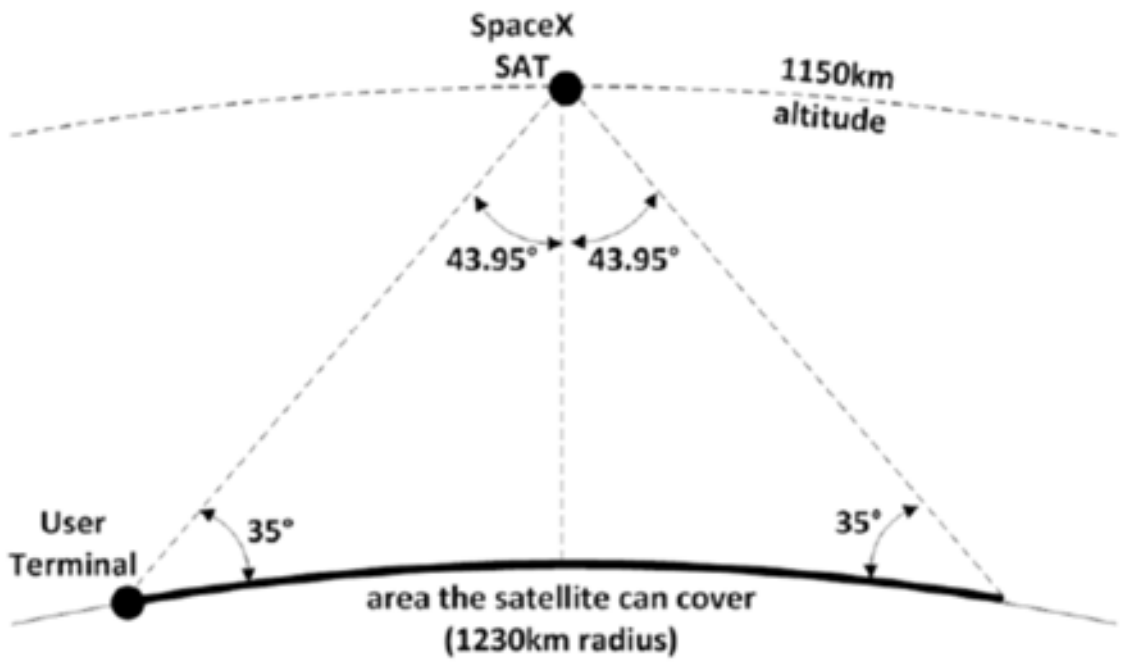

**Figure 4: Steerable Service Range of LEO Beams (1,150 km)**

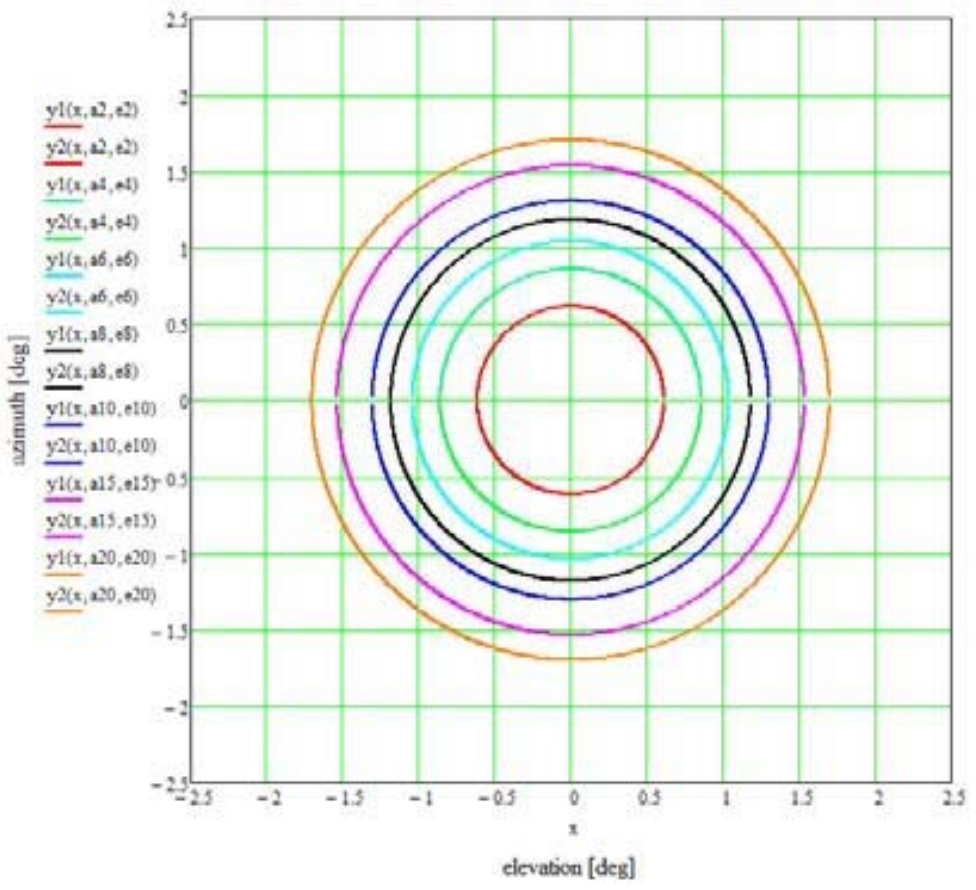

**Figure 6: User (1.5 Degree Beamwidth)**

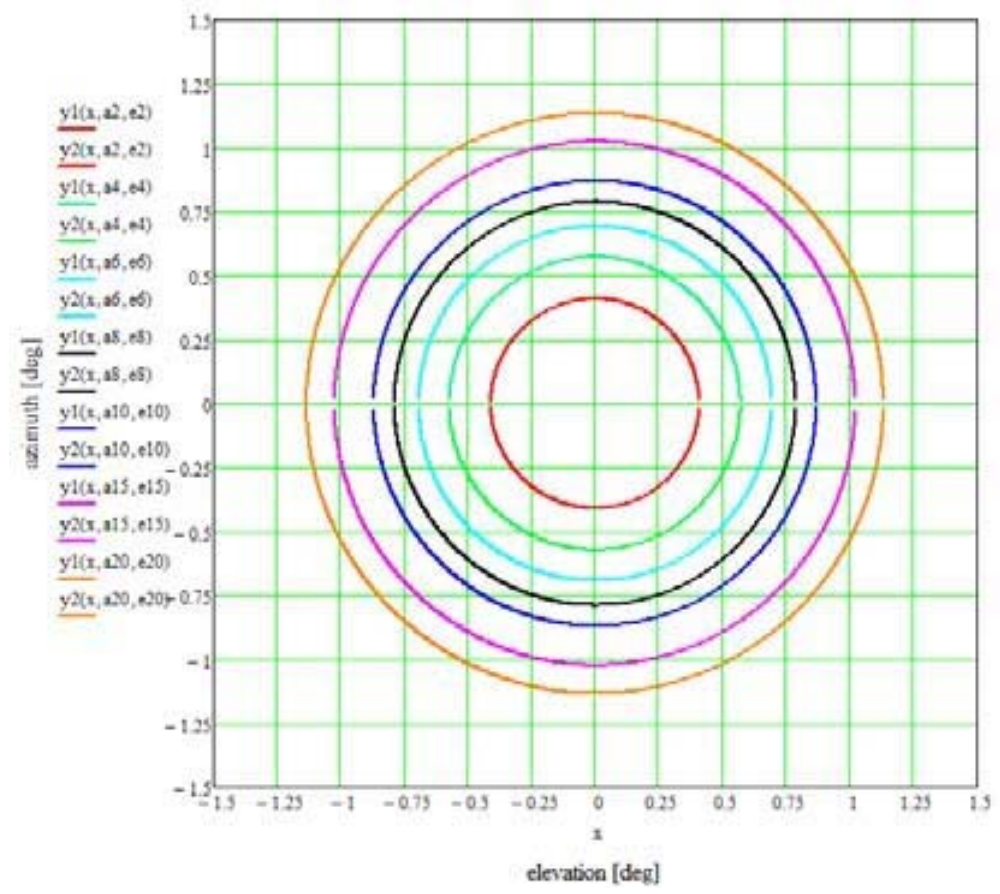

**Figure 7: Gateway (1.0 Degree Beamwidth)**

### RFI case 1 Impact

For RFI case 1, as introduced in section 3.3 emission from transmitters at their allocated frequencies but also in use by a radio telescope, four scenarios can be identified for analysis:

a. Both satellite and radio telescope antennas are pointing at each other. This scenario results in the highest possible RF level at the receiver input of the radio telescope. For a single Starlink satellite transmitting across the full frequency band 37.5 – 42.5 GHz a power level at the input of an ALMA Band 1 receiver of approximately -21 dBm would be measured. This power level being below the damage level of the HEMT based cryogenic amplifiers, estimated to be in the range 0 to +10 dBm, used in the Band 1 receiver. The depicted scenario will be highly unlikely to happen in practice, ALMA antennas should avoid pointing at these satellites. Nevertheless if this scenario occurs strong RFI will be detected by the radio telescope but will not result in any damage.

b. When an ALMA antenna is not pointing at a Starlink satellite while the satellite beam is pointing at ALMA some of its signal can still be received through the antenna sidelobes. As a first approximation it is reasonable to assume that the antenna gain for the sidelobe is at 0 dBi, iow. has the equivalent aperture of a isotropic radiator. For this scenario the received signal at an ALMA antenna would be about 72 dB lower compared to scenario a. Assuming that no saturation in the receiver would occur an additional noise level, expressed as equivalent noise temperature, of about 7 K, for a single satellite, would be detected on top of a system temperature of about 40 K. This scenario can happen in practice but could be avoided by making an arrangement with Starlink agreeing to not point their satellite beams at the ALMA Observatory.

c. When an ALMA antenna is pointing at a Starlink satellite while the satellite beam is not pointing at ALMA a reduced signal can still be received through the sidelobes of the satellite antenna. Unfortunately, little information has been published by Starlink on the antenna properties, we only know that the transmit antennas are so called phased arrays and the main beam properties are given by the patterns in figures 4 and 5 above. Still, this is sufficient information to estimate the satellite side lobe level and calculate the received power at the input of the ALMA receiver. This signal level expressed as equivalent noise temperature, allowing a direct comparison with the system temperature, is on the order of 2,900 K. But like scenario a. ALMA antennas will normally not be pointing at a Starlink satellite.

d. The fourth scenario has both ALMA and satellite antennas not pointing at each other. In this scenario some of the signal transmitted by the satellite will leak to an ALMA antenna through a side lobe. An ALMA antenna will pick this signal up through a side lobe as well. The satellite signal at the input of an ALMA Band 1 antenna would have an equivalent noise temperature of about 0.2 mK. It is emphasized that this is just the interfering signal contribution from one Starlink satellite. When the Starlink Gen 1 constellation is fully deployed nearly 300 satellites will always be above the horizon at the ALMA observatory. Considering that 1) the transmitted signal has random, noise like properties, 2) an ALMA interferometer will not be, delay, tracking satellites, 3) the fast transit time, also in comparison with the ALMA correlator integration / dump times, of these satellites, ~10 – 20 minutes from horizon to horizon, and 4) the random distribution of these satellites across the sky, it is reasonable to assume that the interference from each satellite adds uncorrelated in an ALMA receiver. In other words, the signal power of each satellite can just be added to obtain the total power received from all satellites. Under these assumptions a total noise increase due to the Starlink Gen 1 satellites of about 50 mK at the input of an ALMA Band 1 receiver is expected.

**RFI case 2 Impact**

For RFI case 2, SpaceX has declared in their FCC application that "*To this end, the SpaceX System will comply with the PFD limits and other protections described under the relevant ITU footnotes, resolutions, and recommendations for the protection of RAS*".

This can be understood that they will comply with the RFI thresholds defined by recommendation ITU-R RA.769-2. This recommendation specifies for a frequency of 43 GHz the following thresholds:

- Continuum observations spectral PFD: -227 dBW/($m^2$ · Hz)
- Spectral-line observations spectral PFD: -210 dBW/($m^2$ · Hz)

It is noted that recommendation ITU-R RA.769-2 primarily considers single dish, total power, radio telescopes and does not analyse interference in radio interferometers like ALMA in any detail.

So far Starlink has not provided any further detail how their systems will comply with the ITU recommendations.

### 3.4. Science Implications

From a science perspective, the Starlink transmissions can be considered according to their effects on ALMA continuum and spectral-line observations. In the subsequent discussion we rely heavily on the analysis presented in 'The Science Cases for Building a Band 1 Receiver Suite for ALMA'[23].

When performing continuum observations the telescope processes the largest possible bandwidth, 7.5 GHz currently. This is the same as the frequency range at the top end of Band 1 that will be free from Starlink signal and suggests that the default Band 1 continuum setup could be placed here i.e. between 42.5 and 50 GHz. However, the atmospheric transmission at the high end of the band slowly drops (which will also result in a decline in sensitivity) and users might prefer observing at the bottom end of the band in order to e.g. detect the largest dust grains as foreseen by one of the Level One science cases. A planned future upgrade to an instantaneous bandwidth of 16 GHz (about the same as the entire bandwidth of Band 1) would completely preclude avoiding the Starlink transmissions.

The other Level One science case requires the detection of molecular spectral lines from high-redshift galaxies throughout Band 1. The observed frequency of these relatively narrow spectral features depends on the velocity (redshift) of the emitting source and for the important CO (3-2) transition, the Starlink downlink frequency range corresponds to a redshift range of 7.14-8.22. This is very relevant to studies of the high-redshift universe and cannot therefore be avoided; dealing with the Starlink transmissions is inevitable in this case. The other spectral-line cases are less problematic: none of the important spectral lines listed in Di Francesco et al. fall into the affected region (although SiO at 43.424 GHz is close) and the CCS 4_3-3_2 transition which can be used to observe Zeeman splitting is likewise unaffected (45.379 GHz). This assumes that the lines are observed at their rest frequencies which will be approximately true for all observations of Galactic sources - extragalactic lines with frequencies >42.5 GHz may move into the affected area. Finally, a large number of radio recombination lines (RRL) and a small number of maser lines will fall into the Starlink transmission zone but others (many in the case of the RRL) are available elsewhere in the band.

The effect of the RFI generated by the Starlink transmissions will be to increase the noise in the affected parts of the spectrum, but it should be possible to compensate for this with longer integration times. Where this cannot be done, reductions in sensitivity will have to be accepted although this requires more detailed study. If a significant noise increase is expected, this should be included in the ALMA Sensitivity Calculator so that the correct integration times can be derived.

[23] Francesco et al. (2013).

### 3.5. ALMA Radio Quiet Zone

An important instrument for protecting radio observatories against interference with the help of national authorities is to establish a Radio Quiet Zone (RQZ) around a radio telescope. Such a RQZ agreement foresees coordination of transmitters, even for those operating outside the frequency bands not designated to the RAS, around an observatory. It might even prohibit transmissions at all. What can be agreed for a RQZ agreement is within the jurisdiction of the national authorities who can also enforce possible restrictions.

Currently the Chilean regulatory body, SubTel has an agreement for such a RQZ around ALMA established with the legal entities AUI, for the North American ALMA Partner, and ESO, being the European ALMA Partner. The current RQZ agreement was established in 2013 and will end in 2023. The RQZ agreement only covers ground-based services; air- and space-borne transmitters are specifically excluded. In the future, ALMA partners could investigate whether this RQZ can be expanded to include these additional sources of interference.

### 3.6. Coordination with other RAS stakeholders

Since 2017 ESO is involved with the Committee on Radio Astronomy Frequencies (CRAF), one of the European Science Foundation expert committees, and has currently observer status. On behalf of European radio astronomers, CRAF coordinates activities to keep the frequency bands used by radio astronomy and space sciences free from interference. It works towards this aim by:

- Co-ordinating the case for radio astronomy and space sciences in Europe in discussions with the major public and private telecommunications agencies.
- Acting as the European voice in concert with other groups of radio astronomers in discussions within the international bodies that decide on the use of radio spectrum.
- Initiating and encouraging scientific studies aimed at reducing interference at source and the effects of interference.

ESO also collaborates on this topic with Institut de Radio Astronomie Millimétrique (IRAM) which operates the Plateau de Bure radio telescope, an instrument operating on the same technical principals as ALMA.

Another important stakeholder from the radio astronomy community which ESO keeps close contact with is the Square Kilometre Array Organisation (SKAO). The SKAO is preparing for the construction of the SKA1-Mid radio telescope. This radio telescope will observe in frequency bands in which also satellite constellations will operate.

On a global level ESO is in contact with IUCAF, The Scientific Committee on Frequency Allocations for Radio Astronomy and Space Science. IUCAF operates under the auspices of the International Science Council (ISC). IUCAF is sponsored by the International Astronomical Union (IAU), the International Union of Radio Science (URSI), and the Committee on Space Research (COSPAR).

## 3.7. Spaceborne cloud profiling radar

Although not a vast satellite constellation, single earth observation satellites carrying cloud profiling radar are a threat to radio telescopes to be taken seriously. The very high Effective Isotropic Radiated Power (EIRP) of these radar transmitters can instantly destroy a receiver on a radio telescope.

Already in the early days of ALMA Construction this threat materialized with the launch of CloudSat, a NASA mission. CloudSat is a downward-looking 94 GHz satellite-borne radar, launched in 2005. The peak EIRP of the radar beam is some $4.10^9$ watts, which is sufficient to damage ALMA receivers on the ground if ALMA antennas and the orbiting radar ever look directly at each other. Although the likelihood of this happening is very small, ALMA had to take some operational precautions to avoid receiver damage, and to flag data that is contaminated by radar interference. More information on this issue can be found in ALMA Memo 504 "*The CloudSat Radar and Implications for ALMA*"[24].

At this moment a similar cloud profiling radar mission is being prepared jointly by ESA and JAXA. This cloud profiling radar will be part of the EarthCARE mission to be launched in 2021[25,26]. Also this radar system will be operating at 94 GHz, within ALMA Band 3, like CloudSat. The EIRP will be slightly higher compared to CloudSat due to the increased antenna size of the EarthCARE mounted instrument.

According to the ITU Radio Regulations, being considered an Earth Exploration-Satellite (active) service, cloud profiling radar as mounted on CloudSat and EarthCARE are allowed to operate at 94 GHz. The RAS, in which class ALMA falls, has to accept possible interference from these spaceborne cloud profiling radars. However the ITU RR include an important footnote, 5.562A, to this 94 GHz allocation which reads:

> I*n the bands 94-94.1 GHz and 130-134 GHz, transmissions from space stations of the Earth exploration satellite service (active) that are directed into the main beam of a radio astronomy antenna have the potential to damage some radio astronomy receivers. Space agencies operating the transmitters and the radio astronomy stations concerned should mutually plan their operations so as to avoid such occurrences to the maximum extent possible.*

This mutual planning as mentioned in the footnote has successfully been done for CloudSat in the past but still has to be addressed for EarthCARE. ESO should be in an excellent position to coordinate the planning of the EarthCARE cloud profiling radar.

[24] https://science.nrao.edu/facilities/alma/aboutALMA/Technology/ALMA_Memo_Series/alma504/memo504.pdf

[25] https://earth.esa.int/web/guest/missions/esa-future-missions/earthcare#_56_INSTANCE_v9HD_matmp

[26] https://directory.eoportal.org/web/eoportal/satellite-missions/e/earthcare#sensors

## 4. Actions taken by ESO in support of the Astronomy Community

### Involvement in national working groups

ESO staff are participating in several national working groups convened to address the impacts of satellite constellations, including the Royal Astronomical Society, the French Academie de l'Air et de l'Espace, and the American Astronomical Society (AAS). The European Astronomical Society (EAS) is in the process of establishing a group, with ESO participation.

AAS and NOIRLab organised a “SATCON1” workshop, funded by the US National Science Foundation, which brought together more than 250 scientists, engineers, satellite operators, and other stakeholders from 29 June to 2 July to discuss the impacts of constellations and explore ways to mitigate them (see Walker et al., 2020, for details). A follow-up SATCON2 workshop will focus on regulatory issues in the US context.

### Direct interaction with industry

ESO staff held several calls directly with SpaceX engineers and government relations staff to discuss the issues concerning 4MOST and have interacted with SpaceX, OneWeb and Amazon staff throughout the course of the above working group.

ESO is working on contacts with OneWeb and Amazon, for the purposes of establishing lines of communication in the event that a particular issue impacting ESO arises. In general, however, ESO’s position on industry contacts is that national astronomical societies are best placed to coordinate this interaction, in particular the group set up by the American Astronomical Society (AAS), US, (for SpaceX and Amazon contacts) and to the Royal Astronomical Society (RAS), UK, for OneWeb contacts.

### International awareness raising

In 2017, ESO supported the IAU in submitting a paper to the United Nations Committee on the Peaceful Uses of Outer Space (COPUOS), which raised the issue of Dark Sky protection and recommended that COPOUS request that IAU and the UN Office of Outer Space Affairs (UNOOSA) organise an international conference[27]. COPUOS accepted the recommendation and the Dark and Quiet Skies for Science and Society Workshop, organized by UNOOASA, IAU and the Instituto de Astrofísica de Canarias, was held virtually on 5 - 9 October 2020, with over 900 registered participants.

The conference aimed to raise global awareness of how technological progress, human activities, and artificial illumination degrade the pristine night sky's amateur and scientific investigation. Studies related to different aspects of the impact on observations of dark and quiet skies were presented to provide a broad overview of potential challenges and recommendations in the protection of optical and radio astronomical observatories, the impacts of satellite constellations, protections for dark-sky oases, and light pollution impacts on the bio-environment.

---

[27] The “Dark and quiet skies” proposal as an initiative under the auspices of the Committee on the Peaceful Uses of Outer Space for protecting the environmental observing conditions for large astronomical observatories and world citizens, submitted by the International Astronomical Union (IAU); https://www.unoosa.org/res/oosadoc/data/documents/2017/aac_1052017crp/aac_1052017crp_24_0_html/AC105_2017CRP24E.pdf

On the topic of satellite constellations, ESO led a diverse team of astronomers, industry representatives, and policy and legal experts to study the impact on astronomy and make policy recommendations to observatories, industry, the astronomy community, science funding agencies, national policymakers, regulatory agencies, and international policymakers.

A final workshop report will be released and submitted as a Conference Room Paper during the UNCOPUOS Scientific and Technical Subcommittee, which will take place from 1 to 12 February 2021.

## 5. Council Action

Council is invited to note this document.

**Official statements given by astronomy organisations**

| | |
|---|---|
| European Southern Observatory | https://www.eso.org/public/announcements/ann19029/<br>https://www.eso.org/public/news/eso2004/<br>https://www.eso.org/public/announcements/ann20022/ |
| International Astronomical Union | https://www.iau.org/news/announcements/detail/ann19035/ |
| American Astronomical Society | https://aas.org/media/press-releases/aas-issues-position-statement-satellite-constellations |
| Royal Astronomical Society | https://ras.ac.uk/news-and-press/news/ras-statement-starlink-satellite-constellation |
| Association of Universities for Research in Astronomy | https://www.aura-astronomy.org/news/aura-statement-on-the-starlink-constellation-of-satellites/ |
| Vera C. Rubin Observatory (LSST) | https://www.lsst.org/content/lsst-statement-regarding-increased-deployment-satellite-constellations |
| National Radio Astronomy Observatory | https://public.nrao.edu/news/nrao-statement-commsats/ |
| Square Kilometre Array Organisation | https://www.skatelescope.org/news/ska-statement-on-satellite-constellations/<br>https://www.skatelescope.org/news/skao-satellite-impact-analysis/ |
| Canadian Astronomical Society | https://casca.ca/?p=13575 |
| Spanish Astronomical Society | https://www.sea-astronomia.es/boletin/las-mega-constelaciones-de-satelites-como-amenaza-para-la-observacion-astronomica |